\documentclass{ws-mplb}

\begin{document}

\markboth{Jun Goryo and Mahito Kohmoto}{Adiabatic Process and Chern Numbers}

%
\catchline{}{}{}{}{}
%

\title{Adiabatic Process and Chern Numbers}

\author{\footnotesize Jun Goryo
}

\address{Department of Physics and Mathematics, Aoyama Gakuin University, 5-10-1 Fuchinobe\\
Sagamihara, Kanagawa 229-8558,
Japan
jungoryo@phys.aoyama.ac.jp}

\author{Mahito Kohmoto}

\address{ISSP, University of Tokyo, 5-1-5 Kashiwanoha\\
Kashiwa,  Chiba , 277-8558\\
}

\maketitle

\begin{history}
\received{(Day Month Year)}
\revised{(Day Month Year)}
\end{history}

\begin{abstract}
We study quantum processes with  two or more adiabatic parameters. When the parameters are compactified, a derivative of the Hamiltonian is closely related to a first Chern number. 
This topological formulation is applied to the AC Josephson effect and the spin Hall effect in 
semiconductors.  
\end{abstract}


\section{Introduction}
 
Recently applications of the Chern number   
have attracted a lot of interest. The notable example is the quantum Hall effect. It can be argued that the precise quantization of Hall conductance should be related to topological nature of the systems. In fact the quantization of the Hall conductance in  two-dimensional periodic potentials was proved by using the Kubo formula by Thouless, Kohmoto, Nightingale, and den Nijs\cite{TKNN}. It was shown later by one of us\cite{Kohmoto} that the quantization is due to the topological nature of the problem and Hall conductance is given by Chern numbers in the theory of fiber bundle of differential geometry. The base manifold is T$^2$, which is a magnetic Brillouin zone, and the fiber is wavefunctions. 
Niu, Thouless, and Wu proposed an interesting idea of twisted boundary conditions which yield Chern numbers even in systems without translation symmetry\cite{Niu-Thouless-Wu}. 
Other examples include
quantum Hall effect in  three dimensions\cite{3DQHE},
anomalous quantum Hall effect in ferromagnets\cite{AHE}, 
 spin Hall effect in vortex states \cite{SQHE}, and
Thouless pumping \cite{pumping}.

In this letter we propose a new method involving Chern numbers to have a unified point of view in adiabatic quantum process. 
We shall apply this theory to the examples above. In addition we shall discuss   AC Josephson effect and spin Hall effect in semiconductors\cite{SHE} and obtain a new perspective to these problems.
 
\section{Formulation}
 
We start  with systems with two or more adiabatic parameters. Consider a Hamiltonian $H({\bf{g}}(t))$ with $N(\geq 2)$ adiabatic parameters ${\bf{g}}(t)$.
Schr{\"o}dinger equation is
\begin{equation}
i \frac{d}{d t}  \left |\tilde{\Psi}_{{\bf{g}}(t)}\right>= H( {\bf g}(t))\left|\tilde{\Psi}_{{\bf{g}}(t)}\right>.
\label{schroedinger1}
\end{equation}
In the adiabatic approximation, the parameters {\bf{g}}(t) change in time very slowly.   Then, the solution of Eq. (\ref{schroedinger1}) in the adiabatic approximation is 
\begin{eqnarray}
\left|\tilde\Psi_{n {\bf{g}}(t)}\right>&=&e^{-i \int_0^t d t^{\prime} \epsilon_n({\bf{g}}(t^{\prime}))+ i \gamma_n({\bf{g}}(t))} \left|\Psi_{n {\bf{g}}(t)}\right>, 
\label{adiabatic}
\end{eqnarray}
where  $\left|\Psi_{n {\bf{g}}(t)}\right> $ is an eigenvector of the snapshot Hamiltonian which satisfies Schr{\"o}dinger equation $H({\bf{g}} (t)) \left|\Psi_{n {\bf{g}}(t)}\right>= \epsilon_n({\bf g}(t)) \left|\Psi_{n {\bf{g}}(t)}\right>$. The  presence of a gap around $n$-th level and the absence of the degeneracy 
are assumed.
For a later convenience a phase 
$
\gamma_n({\bf{g}}(t))=i\int_0^t d t^{\prime}  \left<\Psi_{n {\bf{g}}(t^{\prime})} |
\frac{d}{d t^{\prime}} |\Psi_{n {\bf{g}}(t^{\prime})}\right>
$
was introduced.  Note that the  phases of the states can not be determined uniquely at this stage since $\left|\Psi_{n {\bf{g}}(t)}\right>$ is the eigenstates of the instantenious snapshot Hamiltonian. 
When the system makes a loop in the parameter space  we have $H({\bf g}(T)) = H({\bf g}(0))$ and eigenfunctions are same for $t=0$ and $t=T$  except for phase difference. This  difference is called Berry's phase\cite{Berry} and given  by
$
\Gamma_n(l) \equiv \int_0^T dt \gamma_n({\bf{g}} (t))
=i \oint_l d {\bf{g}} \cdot {\bf {A}}^{(n)}_{{\bf g}}, 
$
where
$
{\bf{A}}^{(n)}({\bf{g}})\equiv
\left<\Psi_{n {\bf{g}}}\right|\frac{\partial }{\partial {{\bf{g}}}}\left|\Psi_{n {\bf{g}}}\right> 
$
is a gauge connection in the parameter space \cite{Kohmoto}. 
The phase $\gamma_n({\bf g}(t))$ is nonintegrable, it can not be written as a function just of the end points, because it depends on the geometry of the path connecting them as well. It can be considered as holonomy \cite{eguchi}. Namely, Berry's phase is path dependent and contains geometrical information as a holonomy.

Let us consider derivatives of the Hamiltonian with respect to adiabatic parameters. 
Hereafter, we eliminate the unimportant dynamical phases in Eq. (\ref{adiabatic}) by putting $\epsilon_n({\bf g}(t)) =0$. This manipulation is just to redefine the Hamiltonian as $H({\bf g}(t)) - \epsilon_n({\bf g}(t)) \rightarrow H({\bf g}(t))$. This is allowed because eigenstates are unchanged and 
does not affect the argument below. 
The  expectation values  of the derivative are 
\begin{eqnarray}
&&\left<\tilde\Psi_{n {\bf{g}}(t)}| \frac{\partial H({\bf{g}}(t))}{\partial g_a} |\tilde\Psi_{n {\bf{g}}(t)}\right>
\nonumber\\
&=&\frac{\partial}{\partial g_a} \left\{\left<\tilde\Psi_{n {\bf{g}}(t)}\right| H({\bf{g}}(t))\left|\tilde\Psi_{n {\bf{g}}(t)}\right>\right\}
-\left<\frac{\partial{\tilde\Psi_{n {\bf{g}}(t)}}}{\partial g_a}\left| H({\bf{g}}(t)) \right|\tilde\Psi_{n {\bf{g}}(t)}\right>
\nonumber\\
&&-\left<\tilde\Psi_{n {\bf{g}}(t)}\left| H({\bf{g}}(t)) \right|\frac{\partial{\tilde\Psi_{n {\bf{g}}(t)}}}{\partial g_a}\right>
\nonumber
\end{eqnarray}
\begin{eqnarray}
&=&
\frac{\partial}{\partial g_a} \left\{\left<\tilde\Psi_{n {\bf{g}}(t)}\right| i \frac{d }{d t}\left|\tilde\Psi_{n {\bf{g}}(t)}\right>\right\}
- i \left\{ \left<\frac{\partial \tilde\Psi_{n {\bf{g}}(t)}}{\partial g_a} \right| \left. \frac{d \tilde\Psi_{n {\bf{g}}(t)}}{d t} \right> 
- \left<\frac{d \tilde\Psi_{n {\bf{g}}(t)}}{d t} \right| \left. \frac{\partial \tilde\Psi_{n {\bf{g}}(t)}}{\partial g_a} \right>  \right\}
\nonumber\\
&=&
-i \left\{\left<\frac{\partial \Psi_{n {\bf{g}}(t)}}{\partial g_a} \right| \left. \frac{d \Psi_{n {\bf{g}}(t)}}{d t} \right> 
- \left<\frac{d \Psi_{n {\bf{g}}(t)}}{d t} \right| \left. \frac{\partial \Psi_{n {\bf{g}}(t)}}{\partial g_a} 
\right>\right\}
\nonumber\\
&&+  \dot{\gamma}_n({\bf {g}}(t)) \left\{\left<\frac{\partial \Psi_{n {\bf{g}}(t)}}{\partial g_a} \left|\right. 
 \Psi_{n {\bf{g}}(t)} \right> +  \left<\Psi_{n {\bf{g}}(t)}\left|\right. 
 \frac{\partial \Psi_{n {\bf{g}}(t)}}{\partial g_a}  \right>  \right\}
\nonumber\\
&=&- i \dot{g}_b(t) \left\{  \left<\frac{\partial \Psi_{n {\bf{g}}(t)}}{\partial g_a} \right| \left. \frac{\partial \Psi_{n {\bf{g}}(t)}}{\partial g_b} \right>  -  \left<\frac{\partial \Psi_{n {\bf{g}}(t)}}{\partial g_b} \right| \left. \frac{\partial \Psi_{n {\bf{g}}(t)}}{\partial g_a} 
\right> \right\}
\nonumber\\
&=&- i  \dot{g}_b(t)  \left\{\frac{\partial}{\partial g_a} {A}_b^{(n)}({\bf g}(t)) -  \frac{\partial}{\partial g_b} {A}_a^{(n)}({\bf g}(t))\right\},  
\label{ind-force}
\end{eqnarray}
where $a,b=1,...,N$, repeated indexes are summed over, and the relations 
$
\left<\partial \Psi_{n {\bf{g}}(t)}/{\partial g_a} \left|\right. 
 \Psi_{n {\bf{g}}(t)} \right> =-\left<\Psi_{n {\bf{g}}(t)}\left|\right. 
 \partial \Psi_{n {\bf{g}}(t)}/{\partial g_a}  \right>  
$
and
$
\left| d \Psi_{n {\bf{g}}(t)}/d t \right>=\dot{g}_b(t) 
 \left|\partial \Psi_{n {\bf g}(t)}/{\partial g_b}\right>  
$
are used. Such a term cannot arise in one parameter systems . Note that our formulation is independent of the spatial dimensions or symmetry. 
 
We consider the cases where the parameters are on the $N$ dimentional tori,
$0<g_i\leq 2 \pi$ ($i=1,2,...,N$). 
Consider $N=2$ case. Averaging out the initial values of parameters ${\bf{g}}^{(i)}={\bf{g}}(t=0)$ from Eq. (\ref{ind-force}), we see 
\begin{eqnarray}
\left<\frac{\partial H({\bf{g}})}{\partial g_a}\right>_n&=&\int_{T^2} \frac{d^2 g^{(i)}}{(2 \pi)^2} \left<\tilde\Psi_{n {\bf{g}}(t)}\left| 
 \frac{\partial H({\bf{g}}(t))}{\partial g_a} \right|\tilde\Psi_{n {\bf{g}}(t)}\right>
 \label{exp}
\nonumber\\
&=&{\cal {C}}^{(n)} \epsilon_{ab} \dot{g}_b(t),  
\label{linear-response}
\end{eqnarray}
where $a,b=1,2$, and 
\begin{eqnarray}
{\cal {C}}^{(n)}&=& \int_{T^2} 
\frac{d^2 g}{(2 \pi)^2 i} 
\left[{\nabla}_{{\bf{g}}} \times {\bf{A}}^{(n)}({\bf{g}})\right]_3
=\frac{{\cal{N}}^{(n)}_{Ch}}{2 \pi}. 
\label{Chern}
\end{eqnarray}
Thus it is the first Chern number ${\cal {N}}^{(n)}_{Ch}$\cite{eguchi} and quantized in the unit of $1/2 \pi$.

The Chern number is renowned as the quantized Hall conductance of the Bloch electron in the magnetic field in two dimensions (2D) \cite{TKNN,Kohmoto,Kohmoto_zero}. Here, it should be noted that, the derivation of Eq. (\ref{Chern}) is independent of the real space structure, namely, the periodicity and dimensions. The crucial point to derive Eq. (\ref{Chern}) is the presence of the 
adiabatic parameter ${\bf{g}}(t)$ defined on $T^2$ and the validity of taking the average on ${\bf{g}}^{(i)}$. 
Therefore  Chern numbers have a potential to appear in a wider range in physics. 
    
The condition for nonzero ${\cal {C}}^{(n)}$ is as follows.     
 Introduce "parity" transformation; $H(g_1, g_2) \rightarrow H(g_1, -g_2)$, 
$\left|\Psi_{n(g_1, g_2)}\right> \rightarrow \left|\Psi_{n(g_1, -g_2)}\right>$, 
and "time reversal" transformation;  
$H(g_1, g_2) \rightarrow  H^*(-g_1, -g_2)$,  
$\left|\Psi_{n(g_1, g_2)}\right> \rightarrow  \left|\Psi^*_{n(-g_1, -g_2)}\right>$.   
These are analogous to the parity and time reversal transformations in ${\bf k}$-space of 
2D Bloch electron systems, where ${\bf k}$ is the crystal momentum. 
To ${\cal {C}}^{(n)}$ be nonzero, the system should {\it not} be invariant under these two transformations, since the Chern number counts the vorticity of $\left|\Psi_{n {\bf{g}}}\right>$ in ${\bf{g}}$-space and the vorticity changes its sign under these two transformations \cite{Kohmoto,Kohmoto_zero}.  Then, to preserve the nonzero net vorticity, {\it parity} and {\it time reversal symmetry} have to be broken.  

Consider $N=3$ case. Average out the initial parameters ${\bf{g}}^{(i)} = {\bf{g}}(t=0)$ from Eq. (\ref{ind-force}),  we obtain 
\begin{eqnarray}
\left<\frac{\partial H({\bf g})}{\partial g_a}\right>_n&=& \int_{T^3} \frac{d^3 g^{(i)}}{(2 \pi)^3} \left<\tilde\Psi_{n {\bf {g}}(t)}\left|\frac{\partial H({\bf g}(t))}{\partial g_a}\right|\tilde\Psi_{n{\bf {g}}(t)}\right>
\nonumber\\
&=& {\cal {C}}^{(n)}_{ab} \dot{g}_b(t), 
\label{linear-response3D}
\end{eqnarray}
and
\begin{eqnarray}
{\cal {C}}^{(n)}_{ab}&=&\int_{T^3} \frac{d^3 g}{(2 \pi)^3 i} \left[ 
\frac{\partial A^{(n)}_b ({\bf g})}{\partial g_a} -  \frac{\partial A^{(n)}_a ({\bf g})}{\partial g_b}\right].  
\label{3D-C}
\end{eqnarray}
where $a,b=1,2,3$. Let us introduce a three-vector,
\begin{eqnarray}
{\cal {D}}^{(n)}_c&=&\frac{1}{2} \epsilon_{abc} {\cal {C}}^{(n)}_{ab}
\nonumber\\
&=&\int_{T^3} \frac{d^3 g}{(2 \pi)^3 i} \left[{\bf{\nabla}}_{\bf {g}} \times {\bf A}^{(n)}({\bf{g}}) \right]_c. 
\nonumber\\
&=&\int_0^{2 \pi} \frac{d g_c}{(2 \pi)^2}  {\cal {N}}^{(n)}_c(g_c),  
\end{eqnarray}
where 
\begin{eqnarray}
{{\cal N}_{Ch}}^{(n)}_c(g_c)&=& \int_{T^2(g_c)}\frac{d^2 g}{ 2 \pi i} \left[{\bf \nabla}_{\bf{g}} \times {\bf A}^{(n)}({\bf{g}}) \right]_c 
\end{eqnarray}
is the first Chern number on a two-torus with fixed $g_c$. This is equivalent to Eq. (\ref{Chern}). Because of its topological nature, it is independent of $g_c$.  
Then, we see that 
\begin{eqnarray}
{\cal {C}}^{(n)}_{ab}=\frac{1}{2 \pi} \epsilon_{abc} {{\cal N}_{Ch}}^{(n)}_c.   
\label{3D-Chern}
\end{eqnarray}
An extension to $N>3$ case is straightforward. 

To sum up, in the adiabatic process we have shown definitely the relation between the expectation value of the derivative of the Hamiltonian operator and the Chern number, namely, Eqs. (\ref{linear-response}) and (\ref{Chern}), and Eqs. (\ref{linear-response3D}) and (\ref{3D-Chern}). 

\section{Applications}

We now apply the formulation described above to a number of problems.

\subsection{integer quantum Hall effect in two and three dimensions} 
The electric field is given by ${\bf E}=-\dot{{\bf A}}$.  
The crystal momentum ${\bf k}$ moves adiabatically as
\begin{equation}
{\bf k}(t)={\bf k} - e {\bf E} t=\frac{{\bf g}(t)}{L}.
\label{k-g}
\end{equation}
Then, the electric current carried by a state with ${\bf g}(t)$ is 
\begin{equation}
{\bf J}({\bf g}(t))=-\frac{e}{L^{d-1}} \frac{\partial H({\bf g}(t))}{\partial {\bf g}}, 
\label{current}
\end{equation}
where $d$ denotes the spatial dimensions. 
For a filled $n$-th band in 2D, (\ref{linear-response}) and (\ref{Chern}) give
\begin{eqnarray}
\left<J_i\right>_n=\frac{e^2}{2 \pi} {\cal {N}}^{(n)}_{Ch} \epsilon_{ij} E_j.  
\label{QHE-2D},
\end{eqnarray}
i.e., the quantized Hall current in 2D\cite{TKNN,Kohmoto,Kohmoto_zero}. 

In 3D, we have, from Eqs.  (\ref{linear-response3D}) and (\ref{3D-Chern}),
\begin{eqnarray}
\left<J_i\right>_n=\frac{e^2}{2 \pi L}  \epsilon_{ijk} E_j {{\cal {N}}_{Ch}}^{(n)}_k,
\label{QHE-3D}
\end{eqnarray}
in  agreement  with Refs. \cite{3DQHE}.
These results show that the problem of quantum Hall effects in 2D and 3D can be treated in a unified fashion.

\subsection{Anomalous Hall effect} 
The relations (\ref{k-g}) and (\ref{current}) can be used to this system. 
Therefore, for the filled band, we obtain the quantized Hall current (\ref{QHE-2D}) from Eqs. (\ref{linear-response}) and (\ref{Chern}). From Eqs. (\ref{linear-response3D}) and (\ref{3D-Chern}) it is possible to obtain the 3D current (\ref{QHE-3D}) which is not discussed in Ref. \cite{AHE}.  

\subsection{twisted boundary conditions} 
Niu, Thouless, and Wu pointed out that the Hall conductance of the system with 
twisted boundary conditions (BC) is represented by the Chern number when 
the finite energy gap exists \cite{Niu-Thouless-Wu}. 
Here, we show that the same result can be obtained from Eqs. (\ref{linear-response}) and (\ref{Chern}). 
Compactified parameters $0 \leq (\alpha, \beta) < 2 \pi / L$ (here, $L$ is the size of the system) 
are introduced to represent the twisted BC \cite{Niu-Thouless-Wu}. 
Different values of parameters shows the different BC. 
One can see from Ref. \cite{Niu-Thouless-Wu} that 
the parameters play the role that looks like ${\bf k}$ of the Bloch electron. 
Then, one can put 
\begin{equation}
{\bf g} (t)/ L = (\alpha, \beta) - e {\bf E} t 
\label{twist-parameters}
\end{equation}
(See, Eq. (\ref{k-g})). The electric current carried by a state with ${\bf g}(t)$ is also given by Eq. (\ref{current}). 
Averaging out ${\bf g}(0)=(\alpha L, \beta L)$ is 
justified since it has been shown that the Hall current (conductance) is independent of $(\alpha, \beta)$ 
\cite{Niu-Thouless-Wu}. 
 Then, from Eqs. (\ref{linear-response}) and (\ref{Chern}) the quantized Hall current for the filled $n$-th level (\ref{QHE-2D}) is obtained . The 3D extention is also possible.  

\subsection{Quantum spin Hall effect in a vortex state} 
The quantum spin Hall effect  in a vorex state of 2D superconductor and a superfulid $^3$He film has been investigated \cite{SQHE}. In these systems 
the Bogoliubov quasiparticles are in the Bloch state. The driving force for the spin Hall current 
is not ${\bf E}$ but $\nabla B^z$, where $B^z$ is the magnetic field, and we put 
${\bf g} (t)/ L = {\bf k} + \mu_B \nabla B^z t / 2$, where ${\bf k}$ is the crystal momentum of the quasiparticle and $\mu_B$ the Bohr magneton. The factor 1/2 comes from spin of the quasiparticle.  
The spin current of the $z$-component in the spin space 
carried by the quasiparticle is  analogous to the charged current of Bloch electrons (\ref{current}) \cite{SHE}; 
$$
{\bf J}^z_s=\frac{1}{2L^2} \frac{\partial H^{BdG}_{\bf k}}{\partial {\bf k}}=\frac{1}{2L} \frac{\partial H^{BdG}_{\bf g}}{\partial {\bf {g}}}, 
$$
where $H^{BdG}_{\bf k}$ is the Hamiltonian of the Bogoliubov-de-Gennes equation in the vortex state. 
Then, from Eqs. (\ref{linear-response}) and (\ref{Chern}) we obtain the quantized spin Hall current carried by the filled $n$-th  band 
$$
\left<J^z_{si}\right>_n=\frac{\mu_B^2}{8 \pi} {\cal {N}}^{(n)}_{Ch} \epsilon_{ij} \nabla_j B^z, 
$$ 
which agrees with Ref. \cite{SQHE}.  
 
\subsection{Thouless pumping} Thouless considered 1D Bloch electrons with an adiabatically moving 
periodic potential, which looks ac-like external field at fixed position. He showed that dc charge transfer occurs and its amount per cycle is represented by the Chern number \cite{pumping}. The same result can be obtained from Eqs. (\ref{linear-response}) and (\ref{Chern}). We denote $L$ the spatial periodicity.  
The crystal momentum $k$ varies $0 \leq k < 2 \pi / L$. The Hamiltonian in the $k$-space has a 
form $H_k(x - Vt)$, where $V<<1$ is the velocity of moving potential. 
It is obvious that the system is cyclic in time with a period $T=L / v$. For convenience, 
introduce a parameter $\alpha$ varying $0 \leq \alpha < L$ and consider a Hamiltonian  $H_k(x + \alpha - Vt) \equiv H_{k,\alpha}(t)$ instead of $H_k(x - Vt)$. It is clear that the charge transfer during the period $T$ is independent of $\alpha$. Then, we can average out $\alpha$. 
We may put ${\bf g}(t)=\left(L k, \frac{2 \pi}{L} (\alpha - V t)\right)$ and $H_{k,\alpha}(t)=H({\bf g}(t))$.   
By using Eqs. (\ref{linear-response}) and (\ref{Chern}), and the current (\ref{current}) with $d=1$, we obtain for a  filled $n$-th band  
\begin{eqnarray}
\left<J_x \right>_n=e \frac{{\cal {N}}^{(n)}_{Ch}}{T}.  
\label{j_x}
\end{eqnarray} 
The charge transfer per cycle  agrees to Ref. \cite{pumping}; 
\begin{eqnarray}
\left<\Delta Q\right>_n = \int_0^{T} dt  \left<J_x\right>_n= e {\cal {N}}^{(n)}_{Ch}.   
\label{pumping}
\end{eqnarray}
 It is noted that the above formulation seems to have some relation 
 to the discussions in Ref. \cite{Cohen}. 
 
 



We can clearly see the relation\cite{Avron-Cross} between Thouless pumping and the AC Josephson 
effect by using the general formulation we have introduced. Consider two superconductors 1 and 2 coupled by an insulating film.  
We denote $\theta_1$ and $\theta_2$ the phase of the order parameter of the superconductors 
1 and 2, respectively. The Hamiltonian for the relative phase $\varphi \equiv \theta_1 - \theta_2$ is 
\begin{equation} 
H(t)=-4 E_c \frac{\partial^2}{\partial \varphi^2} - E_J \cos (\varphi - 2 e V t),  
\end{equation}
where $E_c$ and $E_J$ are the charging energy and coupling energy of the junction, 
respectively \cite{deGennes}.  The period of the Hamiltionian $T=\pi / e V$.  We suppose that $V<<1$ and use the adiabatic approximation. 
Since the system is invariant under $\varphi \rightarrow \varphi + 2 \pi$, 
the eigenstates at fixed $t$ are in 
the Bloch states $\Phi^{(n)}_q(\varphi,t)=e^{i q \varphi} U^{(n)}_q (\varphi,t)$, 
where $0 \leq q < 1$ and $U^{(n)}_q (\varphi,t)=U^{(n)}_q (\varphi + 2 \pi,t)$. 
The Hamiltonian for $U^{(n)}_q(\varphi, t)$ is 
\begin{eqnarray}
H_{q,\alpha}(t)=4 E_c\left(-i \frac{\partial }{\partial \varphi} + q \right)^2 - 
E_J \cos (\varphi + \alpha - 2 e V t)
\nonumber
\label{H-q2}
\end{eqnarray}
where $\alpha$ ($0 \leq \alpha < 2 \pi$) is an arbitrary parameter. We can average out $\alpha$ since the change of the relative phase per cycle is independent of $\alpha$.
From Eq. (\ref{H-q2}), we see the one to one correspondence to our formulation.  
We may put 
${\bf g}(t)=(2 \pi q, \alpha - 2 e V t)$ and $H_{q,\alpha}(t)=H({\bf g}(t))$. We see
\begin{eqnarray}
\frac{\partial H({\bf g}(t))}{\partial g_1}=\frac{1}{2 \pi}\frac{\partial H_{q,\alpha}(t)}{\partial q}=\frac{\dot{\varphi}}{2 \pi}. 
\end{eqnarray}
By using Eqs. ({\ref{linear-response})  and (\ref{Chern}), we obtain for the $n$-th level 
\begin{eqnarray}
\left<\dot{\varphi}\right>_n=- 4 \pi eV {\cal {C}}^{(n)}_{\perp}=-2 e V {\cal {N}}^{(n)}_{Ch}.  
\label{phi-dot}
\end{eqnarray} 
The change of relative phase per cycle is    
\begin{eqnarray}
\left<\Delta \varphi\right>_n = \int_0^{T} dt  \left<\dot{\varphi}\right>_n= - 2 \pi  {\cal {N}}^{(n)}_{Ch}.
\label{phi}
\end{eqnarray} 
This result is formally equivalent with the formula for Thouless pumping Eq. (\ref{pumping}). 
In our knowledge, the observable physical measurement for Eq. (\ref{phi}) is not clarified at present.

\section{Spin Hall effect in semiconductors} 
It has been aregued that, in semiconductors with spin-orbit couplings, 
spin current may flow perpendicular to the electric field \cite{SHE}. Then this effect is 
different from the spin Hall effect in a vortex state \cite{SQHE}. 
Here, we would like to consider a condition for a Chern number expression of the spin Hall current in semiconductors by using the general formalism obtained in Section 2.

To discuss the spin Hall effect, we generalize ${\partial H({\bf g})}/{\partial {\bf g}}$  
$$
{\bf v}_{{\cal{O}}}=\frac{1}{2}\left\{\frac{\partial H({\bf g})}{\partial {\bf g}}, {\cal{O}}_{\bf g} \right\}, 
$$
where ${\cal{O}}_{\bf g}$ is an observable and $\{A,B\}=AB+BA$. Let us consider the condition to have  
the Chern number expression for the expectation value of generalized velocity $\left<{\bf v}_{{\cal{O}}}\right>$. When 
\begin{equation}
[H({\bf g}), {\cal{O}}_{\bf g}]=0, 
\label{commutation}
\end{equation}
we have their simultaneous eigenstates, i.e. 
$H({\bf g})\left|\Psi^{(o_{n{\bf g}})}_{n {\bf g}}\right>=0$ \cite{note} and ${\cal{O}}_{\bf g} \left|\Psi^{(o_{n{\bf g}})}_{n {\bf g}}\right>= o_{n \bf g} \left|\Psi^{(o_{n{\bf g}})}_{n {\bf g}}\right>$, 
and if the eigenvalue $o_{n\bf g}$ is independent of ${\bf g}$, i.e., 
\begin{equation}
o_{n{\bf g}}=o_n
\label{eigenvalue}
\end{equation}
we can see that, repeating the derivation of Eqs. (\ref{linear-response}) and (\ref{Chern}) for N=2 case with a slight modification,  
\begin{eqnarray}
\left<{\bf v}_{{\cal{O}}}\right>_n&=&\frac{\epsilon_{ab} \dot{g}_b}{2\pi} \sum_{o_n} o_n {\cal N}_{ch}^{(n,o_n)},  
\end{eqnarray}
where $ {\cal N}_{ch}^{(n,o_n)}$is the Chern number 
\begin{eqnarray}
 {\cal N}_{ch}^{(n,o_n)} &=&\int_{T^2} \frac{d^2g}{2 \pi i} \left[\nabla_{\bf g} \times {\bf A}^{(n,o_n)}_{\bf g}\right]_3, 
\nonumber
\end{eqnarray}
and ${\bf A}^{(n,o_{\bf g})}_{\bf g}=\left<\Psi_{n{\bf g}}^{(o_n)}|\nabla_{\bf g}|\Psi_{n{\bf g}}^{(o_n)}\right>$. Eqs. (\ref{commutation}) and (\ref{eigenvalue}) are 
the conditions for the topological expression of $\left<{\bf v}_{\cal{O}}\right>_n$. The N($\geq 3$)-parameter extension is possible. 

Consider a semiconductor. We apply the relation (\ref{k-g}).  
In case that ${\bf k}$ is not a good quantum number,
we apply the twisted boundary condition technique in Ref. \cite{Niu-Thouless-Wu} 
and use Eq. (\ref{twist-parameters}) instead of Eq. (\ref{k-g}).  
Put ${\cal{O}}_{\bf g}=S_z$ ($S_z$; $z$-component of spin). Then, ${\bf v}_{{\cal{O}}}$ is proportional to the spin current \cite{SHE}. When the spin-orbit coupling is present, $S_z$ does not commute 
with the Hamiltonian and not conserve, i.e. Eq. (\ref{commutation}) is not satisfied.  It is possible to define conserved spins with various manipulations\cite{SHE,spin}, but the eigenvalues of them depend on ${\bf g}$\cite{Qi-Wu-Zhang}, i.e. Eq. (\ref{eigenvalue}) is not satisfied. See, also Ref. \cite{Sheng-Weng-Sheng-Haldane}. We can obtain the topological expression for the spin Hall current in semiconductors, if we succeed to find out a new conserved spin whose eigenvalues satisfy Eq. (\ref{eigenvalue}).


\section*{Acknowledgments}
The authors are grateful to M. Sato, M. Onoda,  and Y.-S. Wu for useful discussions. This work 
is supported by the visiting program of the Institute of Solid State Physics, University of Tokyo in 
2005 and 2006. One of the authors (J.G.) is  supported by Grant-in-Aid for Scientific Research 
for the Japan Society for the Promotion of Science under Grant No. 16740226.


\begin{thebibliography}{99}





\bibitem{TKNN} D. J. Thouless, M. Kohmoto, P. Nightingale, and M. den Nijs, Phys. Rev. Lett. {\bf 49}. 405 (1982). 

\bibitem{Kohmoto} M. Kohmoto, Ann. Phys. (N. Y.) {\bf 160}, 355 (1985). 

\bibitem{Niu-Thouless-Wu} Q. Niu, D. J. Thouless, and Y.-S. Wu, Phys. Rev. B {\bf 31}, 3372 (1985). 

\bibitem{3DQHE} 
B. I. Halperin, Jpn. J. Appl. Phys., Suppl, {\bf 26}, 1913 (1987)  
; G. Montambaux and M. Kohmoto, Phys. Rev. B {\bf 41} 11 417 (1990) 
; M. Kohmoto, B. I. Harpelin, and Y.-S. Wu, {\it ibid.} {\bf 45}, 13 488 (1992). 

\bibitem{AHE} M. Onoda and N. Nagaosa, J. Phys. Soc. Jpn. {\bf 71}, 19 (2002). 


\bibitem{SQHE} A. Vishwanath, Phys. Rev. Lett. {\bf 87}, 217004 (2001) ; 
O. Vafek, A. Melikyan, and Z. Te{\v s}anovi{\' c}, Phys. Rev. B {\bf 64}, 224508 (2001); 
J. Goryo and M. Kohmoto, Phys. Rev. B {\bf 66} 174503 (2002).  

\bibitem{pumping} D. J. Thouless, Phys. Rev. B {\bf 27}, 6083 (1983).  

\bibitem{Cohen} D. Cohen, Phys. Rev. B 68, 155303 (2003). 

\bibitem{Avron-Cross} A somewhat similar relation has been pointed out in; J. B. Avron and M. C. Cross, Phys. Rev. B {\bf 39} 756 (1989). 

\bibitem{SHE} The discussion for the {\it intrinsic} spin Hall effect are given in, S. Murakami, N. Nagaosa, and S. C. Zhang, Science {\bf 301} 1348 (2003); Phys. Rev. B {\bf 69}, 235206 (2004). The {\it extrinsic}  spin Hall effect  is not discussed here. 

\bibitem{Berry} M. Berry, Proc. R. Soc. London {\bf 392}, 45 (1984).  

\bibitem{Kohmoto_zero} M. Kohmoto, Phys. Rev. B {\bf 39} 11943 (1989). 

\bibitem{eguchi} T. Eguchi, P.B. Gilkey, and J. Hanson, Phys. Rep.{\bf 66}, 213 (1980). 
\bibitem{deGennes} P. G. de Gennes, {\it Superconductivity in Metals and Alloys} (Perseus Books, MA, 1999).  




\bibitem{note} We should remind the readers that we put $\epsilon_n({\bf g}(t))=0$ to eliminate the dynamical phase factor in Eq. (\ref{adiabatic}). 

\bibitem{spin} J. Shi, P. Zhang, D. Xiao, and Q. Niu, Phys. Rev. Let. {\bf 96} 076604 (2006). 

\bibitem{Qi-Wu-Zhang} An exceptional case is discussed in 2D. See,  
X.-L. Qi, Y.-S. Wu and S. C. Zhang, cond-mat/0505308.  

\bibitem{Sheng-Weng-Sheng-Haldane} D. N. Sheng, Z. Y. Weng, L. Sheng, and F. D. M. Haldane, cond-mat/0603054. 




\end{thebibliography}
\end{document}